\documentstyle[aps,multicol,epsf,psfig]{revtex}

\begin{document}
\draft

\title{Modeling relaxation and jamming in granular media}
\author{B.~Kahng$^{1,2}$, I.~Albert$^1$, P.~Schiffer$^3$, and A.-L.~Barab\'asi$^1$} 
\address{$^1$ Department of Physics, University of Notre Dame, 
Notre Dame, IN 46556\\
$^2$ Department of Physics and Center for Advanced 
Materials and Devices, Konkuk University, Seoul 143-701, Korea\\
$^3$ Department of Physics, Pennsylvania State University, University Park, PA  16802
}
\date{\today}
\maketitle 

\thispagestyle{empty}
\begin{abstract} 
We introduce a stochastic microscopic model to investigate the jamming 
and reorganization of grains induced by an object moving through a
granular medium. The model reproduces the experimentally observed 
periodic sawtooth fluctuations in the jamming force 
and predicts the period and the power spectrum in terms of the controllable
physical parameters. It also predicts 
that the avalanche sizes, defined as the number of displaced
grains during a single advance of the object, follow a power-law,
$P(s)\sim s^{-\tau}$, where the exponent is 
independent of the physical parameters.
\end{abstract} 
\begin{multicols}{2}

Granular materials are composed of many solid particles that  
interact only through contact forces, displaying a variety of 
behavior that distinguishes them from other forms of matter
such as liquids or solids. While granular materials can flow like 
a liquid, unlike liquids, they reach a jammed state 
when stressed\cite{jamming}. In the jammed state, 
which has analogies in a variety of physical systems such as 
dense colloidal suspensions, traffic flows, and spin 
glasses\cite{nagel,glass}, 
the local dynamics of grains is frustrated by close contacts 
between neighboring grains. 
Although jamming in granular materials has previously been 
discussed in the context of the gravitational stress 
induced by the weight of the grains,
it can be induced by any compressive stress, such as the stress
generated by an object traveling through a granular material. 
Indeed, there is experimental evidence \cite{reka,albert} 
that a solid object being pulled slowly through a granular 
medium is resisted by local jamming, and can only advance 
with large-scale reorganizations of the grains. The jamming 
and reorganization phenomenon, which can be detected through 
the drag force acting on the object in  granular medium,
reflects both the nature of
stress propagation through the medium and the dynamics of 
the granular medium \cite{jaeger,liu,miller,jia,kolb,copper,tk,ANgadi,vanel,sawtooth}. 
The drag force opposing the motion 
of the object originates in the force needed to induce such 
reorganizations and exhibits strong fluctuations with
a stick-slip
character associated with the reorganization of the grains
\cite{albert}.
While these phenomena were documented extensively experimentally, 
a theoretical microscopic study of the behavior of jamming 
and relaxation of the drag force has never been attempted. 

In this paper we introduce a microscopic model for the motion of a 
vertical cylinder through a granular medium, describing the jamming and 
reorganization of grains in terms of 
compression and relaxation of elastic springs\cite{scalar}
with random thresholds. 
The model can reproduce many of the experimentally documented features
of the drag force, such as the sawtooth shape of the temporal behavior
and the main features of the power spectrum. It predicts
the period of the sawtooth pattern in terms of physical parameters
such as depth, diameter, and cylinder velocity 
in agreement with
recent experimental results \cite{albert,reka,new}.
In the jammed state, it predicts an avalanche-like 
relaxation of the grains, the avalanche size following a 
power-law distribution.
The model also predicts that the critical relaxations 
generating the sudden drops are nucleated preferentially from the bottom part 
of the cylinder. 
Finally, we find that the temporal pattern of the drag force
can change depending on the elasticity of the grain medium:
when grains are rigid 
the temporal pattern is periodic and the avalanches follow a power-law, 
however, when they are more elastic, the temporal pattern is random 
and the avalanche size decays exponentially.

{\it Microscopic model ---} The model was constructed to 
emulate the drag experienced by a vertical cylinder
of diameter $d$ inserted to a depth $H$ in a granular bed \cite{reka,albert}.
The grains move with constant speed $v$ in 
the positive $x$ direction, 
pushing the cylinder in the same direction. The motion of the cylinder
is constrained by a fixed stop which is coupled to the cylinder
through an external spring with spring constant $K$. The force on the fixed 
stop is equivalent to the drag force, $F(t)$, on the cylinder 
as a function of time. We refer to the spring located between 
the cylinder and fixed stop as the external spring. As the granular medium moves in
the positive $x$ direction, the grains in contact with the cylinder's surface push
the cylinder with small forces whose sum is 
the total drag force.
To model the heterogeneous nature of granular drag 
we regard the surface of the cylinder as a planar rectangle 
partitioned into $d\times H$ cells of unit size.

Grains push the cylinder only if they are compressed, and we thus 
model each cell as a "grain-spring". 
The magnitude of the force $f_{i,j}(t)$ exercised by the spring in cell
$(i,j) \quad (i=1..d, \quad j=1..H)$ is given by Hooke's law,
$f_{i,j}=f_{i,j}(0)+k_{i,j}x_{i,j}$ where $f_{i,j}(0)$ is the $x$ component 
of the ambient force,
$k_{i,j}$ is the grain spring constant, and $x_{i,j}$ is the deviation 
of the spring's length from its uncompressed value. 
It is well known that the pressure in granular media increases 
linearly with depth and  will increase the ambient force $f_{i,j}(0)=f_0j$ \cite{grain_note}.
The grain-spring constant should be interpreted as describing 
the elasticity of the force chains instead of the individual grains. 
The force chains are expected to get stiffer with depth, since the participating 
grains are more compressed, allowing less room for configurational changes.
Thus we expect that the spring constant $k_{i,j}$ will also increase linearly, 
$k_{i,j}=k_0j$.

If the grains are too compressed, they will 
fail by slipping relative to the cylinder' surface and each other, thus relaxing 
the local forces. As a result, the total force acting 
on the cylinder decreases, and the cylinder slips relative 
to the grains. To model this microscopic failure  we 
introduce a critical threshold $g_{i,j}$, which is a random 
variable uniformly distributed between $[g_0 j, g_1 j]$, where 
$g_0$ and $g_1$ are constants. When the elastic force on a 
grain spring exceeds its critical threshold, ($f_{i,j} \ge g_{i,j}$), the spring is 
relaxed to its equilibrium position, $f_{i,j}=f_{i,j}(0)$, 
and the threshold $g_{i,j}$ is newly updated by a new random number.  

As grains advance in the positive $x$-direction 
by the distance $v\delta t$ during time interval $\delta t$, they increase the total force 
acting on the cylinder, compressing the external spring $K$ as well. The balance between
the cumulative action of the grain springs and the opposing force of the external spring
allows the cylinder to move in the $+x$ direction by a distance $\ell$, 
\begin{equation}
\ell = k_t v {\delta t}/(k_t +K),
\end{equation} 
where  $k_t =k_0 d H(H+1)/2,$
is the collective spring constant of the 
grain springs. The distance (1) was obtained by balancing the 
collective elastic forces of the grain springs and the external spring
on each side of the cylinder, 
\begin{equation}
\sum_{i,j} k_{i,j} \delta x_{i,j}=K \ell, 
\end{equation}
where $\delta x_{i,j}=v \delta t -\ell$. 
After obtaining $\ell$, the effective compression of grain springs 
can be determined from $\delta x_{i,j}=K v \delta t /(k_t +K)$, leading 
to an increase in the grain spring force by   
$\delta f_{i,j}= k_0 j K v {\delta t}/(k_t +K).$
When the grain springs are much stiffer than the external springs 
($k_t \gg K$), corresponding to the case in which the experiment
was performed, the increased grain spring force acting on the 
cylinder at $(i,j)$ becomes 
\begin{equation}
\delta f_{i,j}= k_0 j K v {\delta t}/k_t.  
\end{equation}
The situation suddenly changes if a grain slips, i.e. the force 
$f_{i,j}$ on a grain spring reaches its threshold $g_{i,j}$. 
We reset the force to $f_{i,j}(0)$,
and the threshold is updated \cite{note}. 
After this update, the balance between the elastic
forces on each side of the cylinder breaks down, because the total force
acting from the grains on the cylinder has dropped by 
$f_{i,j}(t)-f_{i,j}(0)$. As a result the cylinder will move
backward (in the negative $x$ direction), pushed by the external spring,
compressing further the remaining grain springs.
The displacement of the cylinder can be calculated by using 
the balance equation (2), where the newly updated sites (i.e., those having 
$f_{i,j}=f_{i,j}(0)$) are excluded from the summation.

There are two possible outcomes of this slip event. 
First, if this sudden compression of all grain 
springs will not cause any more springs to reach their thresholds, after establishing 
a new equilibrium we continue the continuous compression of 
all springs by the motion of the grains with velocity $v$.
However, in some cases the discontinuous
increase of the force on the grain springs will cause some other 
springs to reach their thresholds. In this case the updating (replacing each broken 
spring with an uncompressed one and calculating the new equilibrium) is repeated 
until no further reorganizations occur.
The time is then incremented, followed by the advance of all grain springs,
leading to a repetition of the above processes through
compression and new updating.
The dynamics of the model are similar to the random fuse model 
in one dimension \cite{fuse} describing fracture of a fiber bundle in the sense that
when one spring (bond) breaks down, the load from it is shared by
others. There is a significant difference, however, in that within the current model 
the springs can be re-compressed, resulting in a stationary 
dynamics, while in the random fuse model a bond is permanently disconnected 
once it has burned out.

The stochastic model described above offers a microscopic 
description of the system investigated experimentally. Despite its simplicity, 
as we show next, it accounts for many key factors of
the observed behavior, and offers insight
and quantitative predictions that were not available experimentally \cite{reka,albert}.

{\it Sawtooth pattern ---} 
A characteristic feature of the drag force observed in the experiments 
is that the force on a cylinder, $F(t)$, increases linearly, followed by a a sudden drop in $F(t)$,
corresponding to a collective failure and reorganization of the grains.
As Fig. 2a-c shows, this sawtooth pattern is fully reproduced by the model.
The linear increase corresponds to a continuous compression of both the grain 
springs and the external spring. At a certain point, however,  a grain spring fails,
which results in a collective and subsequent failure of all other
springs in the system, since they are compressed to near their thresholds.
Thus the stick-slip motion observed in the experiments correspond to two 
regimes: in the linear regime we see a linear convergence to the critical state, 
where all the springs are more or less simultaneously compressed towards 
their critical threshold. The sudden drop
corresponds to an avalanche like spreading of a failure 
as soon as the critical or fragile state has been reached. The advantage 
of the presented model is that it allows us to quantitatively characterize 
the resulting stick-slip process. 

{\it Linear Regime ---} What does
determine the slope of the $F(t)$ signal in the linear regime?
Eq. (4) predicts that the 
drag force, $F(t)=\sum_{i,j} f_{i,j}(t)$, increases linearly with
time with the slope $\frac{1}{v}\frac{dF}{dt} \sim K$ in the jammed state. This linear
increase is in complete agreement with the experimental
results (see Fig. 2 of \cite{albert}).
Furthermore we predict that the slope is independent of the experimental details,
but depends only on the spring constant of the external spring, which is again 
consistent with the experimental findings. 

{\it Failure and depth dependence ---} When updating occurs 
over the entire system, the drag force
drops suddenly, because most grain springs are reset to 
their equilibrium positions, and $f_{i,j}(t)=f_{i,j}(0)$.
Accordingly, we expect the drag force to exhibit a sawtooth pattern. This
is supported by extensive numerical simulations whose  results are summarized 
in Fig.2a-c. This allows us to determine the average value of the drag force, 
that has been investigated extensively both experimentally and numerically (see Ref. [5]).
Since the force at $(i,j)$ is independent 
of $i$ and proportional to $j$, we find that the average drag force $\bar F$ 
over time is proportional to $\sim d H^2$. This result is confirmed by numerical 
simulation as well (see Fig.2d), and is in agreement with the experimental 
results \cite{albert}. 

{\it Power spectrum ---}
To better characterize the fluctuations in the system, we 
measured the power spectrum of the time dependent drag force 
as predicted by the model. In \cite{albert} it has been found that 
the power spectrum is characterized by a few prominent peaks and sub-harmonics, 
determining the period of the signal, followed by a power-law tail at high frequencies 
which decays as $f^{-2}$. The numerically determined power spectrum has 
the same features (Fig 2e), exhibiting an  $\sim 1/f^2$
behavior at high frequencies and peaks at low frequencies.
The origin  of the $1/f^2$ behavior comes from fluctuations of critical relaxation 
time due to random thresholds. The position of the largest peak corresponds
to the inverse of the period of the sawtooth pattern.
Thus, estimating the peak position, we can determine ($1/T$), the period
of the sawtooth signal (see the inset of Fig.2e).
Furthermore, this period can be predicted analytically as follows: the
critical updating occurs when $\delta f_{i,j}$ is increased
to the maximum value of threshold $\sim j g_1$.
The time required to reach this critical force through jamming is
\begin{equation}
T \sim k_t (g_1-g_0)/(k_0 K v) \sim (g_1-g_0)H^2 d / K v, 
\end{equation}
which represents the period. The numerical simulation data 
for different depths, diameters, and elastic spring constants $K$ (shown
in the inset of Fig.2e) confirm Eq.(4), and are also in agreement with 
recent experimental results\cite{new}.

{\it Avalanches ---} A quantity that has not been measured experimentally, but
can be determined in the present model, is the avalanche size distribution $P(s)$. 
When a collective failure occurs, this will result in the simultaneous 
failure of a certain number of grains springs (but not all) creating an 
avalanche of failures.
The avalanche size, $s$, is defined as the number of springs newly updated
in a single advance of the cylinder. We find that the avalanche
size follows a power-law distribution, $P(s)\sim s^{-\tau}$
with $\tau \approx 2.4(1)$ (see Fig. 3a). Since  power-law 
distribution of events are common only at the critical point 
of spatially extended systems indicating that the 
continuous compression of the grains brings the system close 
to a critical state. Furthermore 
we find that the exponent $\tau$ is universal, independent of physical 
parameters such as depth $H$, width $d$, spring constant $K$,
grain spring constant $k_0$, and threshold $g_1-g_0$.
Note that while the avalanche sizes could not be measured experimentally,
bulk imaging techniques such as MRI, allowing one to compare the grain positions 
before and after an avalanche, could offer a quantitative check of our predictions. 

The simulations indicate that the critical avalanches are
nucleated preferentially near the bottom of the cylinder where 
the stress is largest (see inset to Fig. 3a). 
When a spring at large $H$ is relaxed suddenly, the load taken over by the other springs 
is  commensurately large, and has a higher probablity of nucleating a
large-scale reorganization 
of the grains. In contrast, breakdown of springs at small $H$ is less likely 
to nucleate avalanches.

Finally, the model predicts that the elastic properties of the granular 
media have a strong impact on the temporal characteristics of the drag force.
When grains are sufficiently
soft, $k_t \ll K$, the drag force develops a random signal rather than sawtooth,
(inset to Fig. 3b). Such random characteristics also occur when the depth $H$ is small
enough to satisfy the condition $k_t \ll K$ for a given grain 
spring constant $k_0$.  
Interestingly, in this case, the avalanche size distribution does
not follow a power-law, but decays exponentially 
as shown in Fig. 3b.
  
In summary, we have introduced a stochastic model that describes
the jamming  and reorganization of grains associated with dragging an object 
through a granular medium.
The model reproduces the sawtooth pattern of temporal evolution of the 
drag force  and the $1/f^2$ high frequency tail of the power 
spectrum, and predicts a power-law avalanche size distribution. This 
excellent agreement with the experiments is surprising because the 
model offers a mean-field treatment of the force chains which are known to be
the basic mechanism of the stress propagation through grains. Indeed, 
the force chains are only included implicitly through the depth dependence
of the grain-spring elastic constant.
Thus improvements based on a more detailed handling of the force chains 
could be envisioned, but
our model  offers a crucial starting point for a detailed 
understanding of motion through granular media, 
and it offers a basis for more realistic modeling efforts.

This work is supported by the Petroleum Research Fund  
administered by the ACS, the Alfred P. Sloan Foundation, 
NSF Grants No. PHYS95-31383 and No. DMR97-01998, NASA grant 
number NAG3-2384 and by grants No. 2000-2-11200-002-3 from 
the BRP program of the KOSEF.

\end{multicols}

\newpage
\begin{figure}
\centerline{\psfig{figure=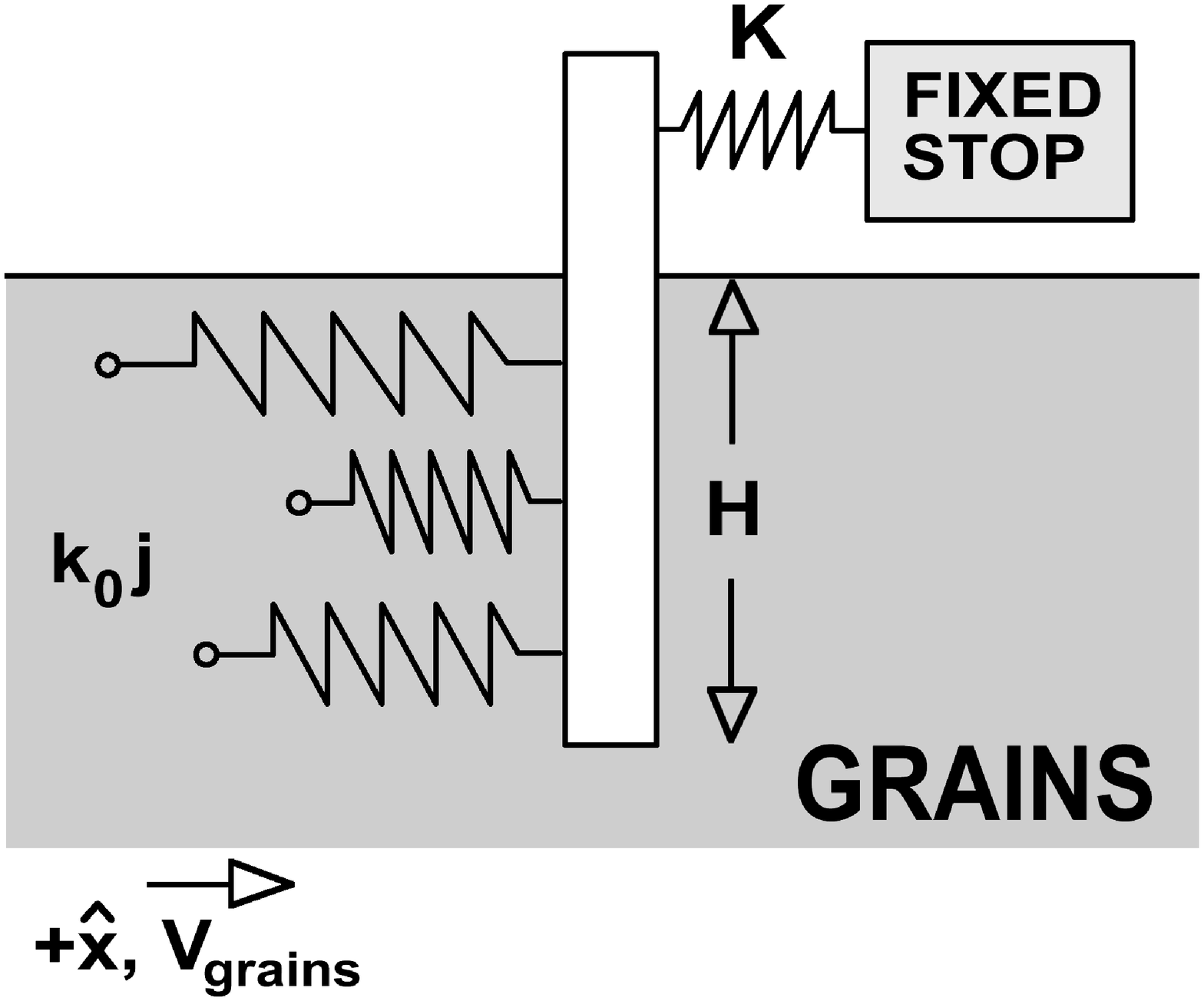,width=8.5cm,angle=-90}}
\caption{Schematic illustration of the stochastic spring model. 
The shaded area indicates the granular medium, which moves with velocity 
$v$ in the positive $x$ direction. The motion of the cylinder that tries to move along with the grains is opposed
by a fixed stop coupled to the cylinder through a spring with spring constant $K$. 
We model the grains opposing the movement as springs with spring 
constant $k_0j$.} 
\end{figure}
\newpage
\begin{figure}
\centerline{\epsfxsize=7.0cm \epsfbox{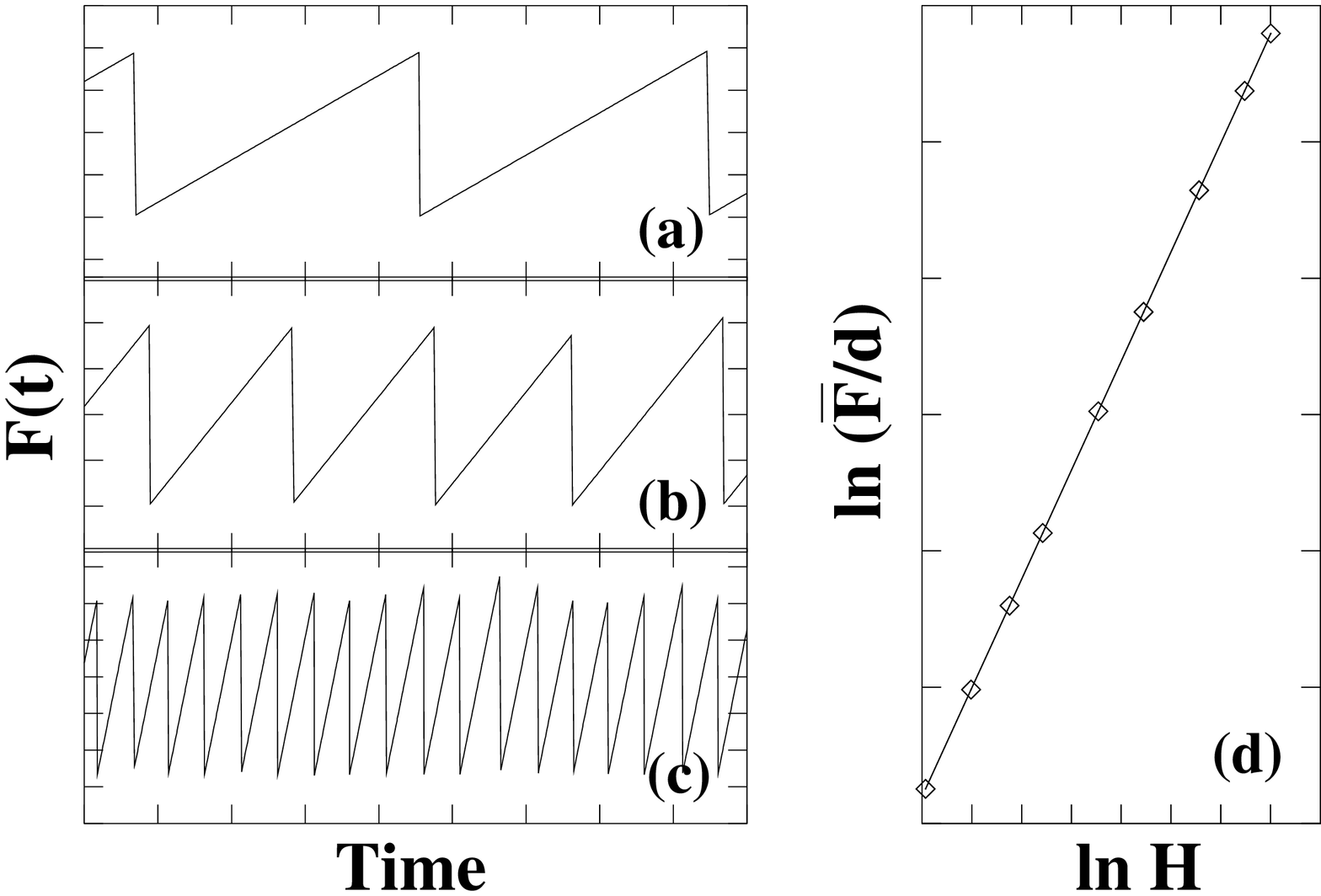}}
\centerline{\epsfxsize=7.0cm \epsfbox{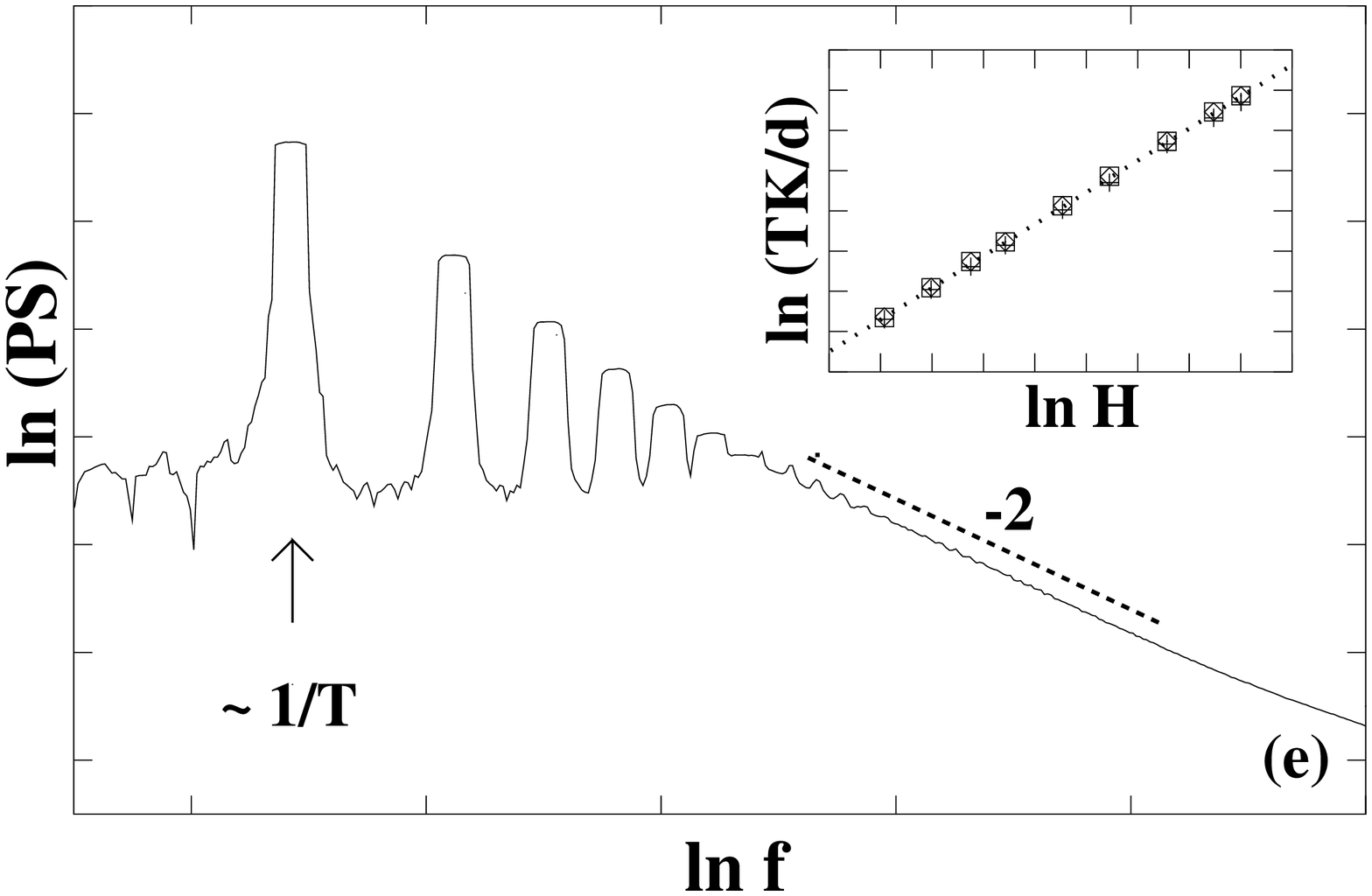}}
\caption{(a-c) Plot of the drag force $F(t)$ as a function of time 
for $d=2$, $H=1000$ (a), $d=1$, $H=1000$ (b), and $d=1$, $H=500$ (c), 
where $f_0=0.5$, $g_0=0.5$, $g_1=0.7$, $K=1$, $k_0=1$, and $v=1$ 
are used. 
(d) Double logarithmic plot of ${\bar F}/d$ versus $H$ for different 
diameters $d=1$ and 5. 
The solid line with slope $2.0$ is obtained by a least square fit. 
(e) Double logarithmic plot of the power spectrum (PS) versus 
frequency. 
Inset: Double logarithmic plot of $TK/d$ versus depth $H$
for different spring constants $K=1$ and $K=5$ and cylinder diameters 
$d=1$ and $d=5$ of the cylinder. The data are well collapsed 
on the dotted line with slope $1.98$, obtained by a least square fit, 
predicted by Eq.(4).}   
\end{figure}
\newpage
\begin{figure}
\centerline{\psfig{figure=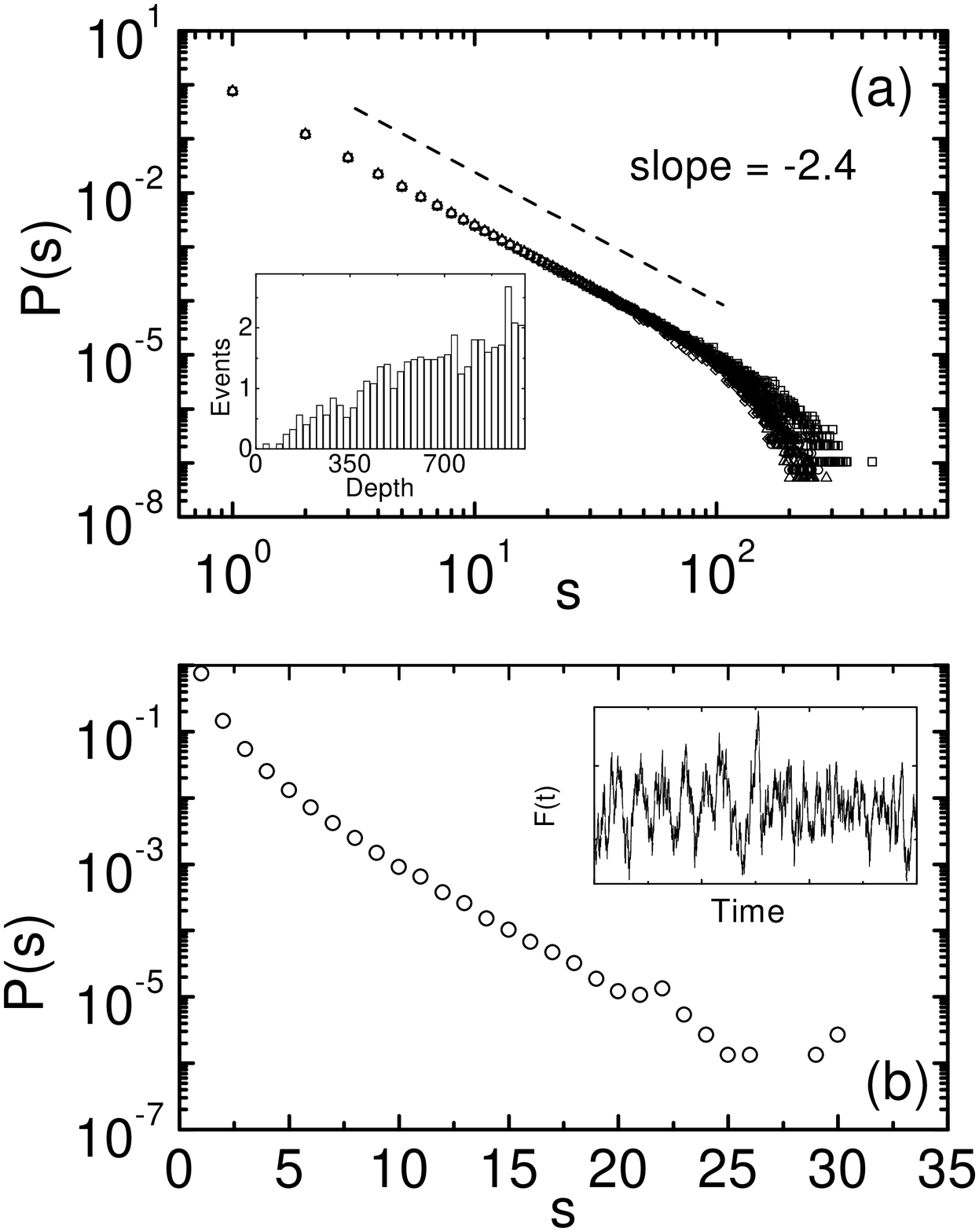,width=8.5cm}}
\caption{a)  Logarithmic plot of the avalanche size distribution $P(s)$ versus
size $s$ for different
diameters $d=1$ and $2$, depths $H=1000$ and $2000$,
elastic spring constants $K=1$ and $10$,
and grain spring constants $k_0=1$ and $10$, where 
critical avalanches which 
contribute an isolated point are excluded in the accumulation. All data, averaged 
over 5000 configurations, are collapsed to $P(s)\sim s^{-2.4}$.
The inset shows the number of avalanches nucleated at a certain depth
averaged over a 25 point depth interval.
b)  Semi-logarithmic plot of the avalanche size distribution versus size for
$k_0=10^{-6}$ showing an exponential distribution. 
The data are averaged over 5000 runs.
The inset shows a plot of the drag force as a function of time 
under the same condition used in Fig.2b, but $k_0=10^{-6}$.} 
\end{figure}

\begin{thebibliography}{99}
\bibitem{note}{Note that we choose $g_0 > f_0$
so that the initial stress $f_0j$ is always smaller than threshold,
$g_0 j$.}
\bibitem{grain_note}{
Note that the depth we consider here is shallower than the 
Janssen's critical depth, thus the average pressure increases linearly for
all considered depths.
}
\end{thebibliography}
\end{document}